\documentclass[preprintnumbers,aps,pra,groupedaddress]{revtex4}

\begin{document}
\draft
\title{Antineutrinos produced from $\beta$ decays of neutrons cannot be in
coherent superpositions of different mass eigenstates}
\author{Shi-Biao Zheng}
\thanks{E-mail: t96034@fzu.edu.cn}
\address{College of Physics and Information Engineering, Fuzhou University,\\
Fuzhou 350108, China}

\begin{abstract}
The entire wavefunction of the antineutrino-proton-electron system, produced
by the $\beta $ decay of a neutron is analyzed. It is proven that the
antineutrino cannot be in coherent superpositions of different mass
eigenstates, irrespective of the initial momentum distribution of the
neutron.
\end{abstract}

\vskip0.5cm

\narrowtext

\maketitle

\bigskip The discovery of neutrino oscillations represents one of the most
remarkable physical advancements achieved during the past three decades
[1-6]. These phenomena cannot be interpreted within the framework of the
standard model [7]. According to the standard model, there are three
generations of leptons, each composed of a charged lepton and a neutrino.
Charged leptons gain their mass by the Yukawa interaction, where the
left-handed and right-handed leptons are coupled by the Higss scalar field.
Each lepton generation has its own flavor. Charged leptons with different
flavors have very different masses, e.g, the muon is heavier than the
electron by two orders of magnitude. In distinct contrast, all neutrinos are
left-handed so that they cannot get mass from the Higss field. The leptons
belonging to the same family can be transformed into each other by
exchanging a W boson. However, the Lagrangian of the lepton model does not
contain any term that can change the flavor of a lepton.

To interpret neutrino oscillations, it was postulated that each neutrino
flavor corresponds to a linear superposition of three distinct mass
eigenstates [8], 
\begin{equation}
\left\vert \nu _{\alpha }\right\rangle =%
\mathop{\displaystyle\sum}%
\limits_{j}U_{\alpha j}\left\vert \nu _{j}\right\rangle ,
\end{equation}%
where $\alpha =e,\mu ,\tau $ denotes the neutrino flavor, while $j=1$ to $3$
labels the mass eigenstate. When the momentum of the neutrino ($p$) is much
larger than the mass $m_{j}$, the energy of the mass component $\left\vert
\nu _{j}\right\rangle $ can be well approximated by $E_{j}\simeq
p+m_{j}^{2}/(2p)$. After a propagation time $t$, $\left\vert \nu
_{j}\right\rangle $ accumulates a phase $\phi _{j}=-E_{j}t$. Consequently,
the initial flavor eigenstate $\left\vert \nu _{\alpha }\right\rangle $
evolves to $%
\mathop{\displaystyle\sum}%
\limits_{j}U_{\alpha j}e^{i\phi _{j}}\left\vert \nu _{j}\right\rangle $. The
probability for the neutrino to remain in the flavor eigenstate $\left\vert
\nu _{\alpha }\right\rangle $ is given by 
\begin{equation}
P_{\left\vert \nu _{\alpha }\right\rangle \rightarrow \left\vert \nu
_{\alpha }\right\rangle }=\left\vert 
\mathop{\displaystyle\sum}%
\limits_{j}U_{\alpha j}U_{\alpha j}^{\ast }e^{i\phi _{j}}\right\vert ^{2}.
\end{equation}%
Due to the phase difference accumulated by different neutrino mass
eigenstates, $P_{\left\vert \nu _{\alpha }\right\rangle \rightarrow
\left\vert \nu _{\alpha }\right\rangle }$ exhibits oscillatory behaviors.
The quantum coherence between the mass eigenstates is responsible for these
behaviors.

This interpretation is valid only when the neutrino can be in the
superposition state of Eq. (1) at the production. Previously, it was
realized that the neutrino emitted by an unstable particle with a definite
momentum is necessarily entangled with the particles accompanying the
neutrino [9-11]. This entanglement would destroy the coherence between the
neutrino's mass eigenstates. To overcome this problem, it was argued that
the neutrino can be disentangled with the accompanying particles when the
momentum uncertainty of the unstable particle is sufficiently large [9-14].
However, I find that different mass eigenstates of such a neutrino, if they
exist, are necessarily correlated with different joint momentum states of
the entire system, including the neutrino and the accompanying particles,
which prohibits occurrence of interference effects between the mass
eigenstates. I will illustrate this point with the $\beta $ decay of a
neutron.

{\sl Lemma} {\sl 1:} For the $\beta $ decay, when there is no mass-momentum
entanglement, the produced electron antineutrino has a definite mass.

The wavefunction of the neutron, before undergoing the $\beta $ decay, can
be expanded as%
\begin{equation}
\left\vert \psi _{n}\right\rangle =\int \varphi ({\bf P}_{n})d^{3}{\bf P}%
_{n}\left\vert {\bf P}_{n}\right\rangle ,
\end{equation}%
where $\left\vert {\bf P}_{n}\right\rangle $ denotes the momentum eigenstate
of the neutron with the eigenvalue ${\bf P}_{n}$. Suppose that the electron
antineutrino produced by the $\beta $ decay possesses three different mass
eigenstates, which are not entangled with different momentum eigenstates of
the antineutrino, proton, and electron. Then the state of the composite
system can be written as%
\begin{equation}
\left\vert \psi _{\nu +p+e}\right\rangle =\int d^{3}{\bf P}_{\nu }d^{3}{\bf P%
}_{p}d^{3}{\bf P}_{e}F({\bf P}_{\nu },{\bf P}_{p},{\bf P}_{e})\left\vert 
{\bf P}_{\nu },{\bf P}_{p},{\bf P}_{e}\right\rangle (%
\mathop{\displaystyle\sum}%
\limits_{j}C_{j}\left\vert \stackrel{-}{\nu }_{j}\right\rangle ).
\end{equation}%
Here $\left\vert {\bf P}_{\nu },{\bf P}_{p},{\bf P}_{e}\right\rangle $
represents joint momentum eigenstate of the entire system, where the
antineutrino, proton, and electron are all in their momentum eigenstates
with the eigenvalues ${\bf P}_{\nu }$, ${\bf P}_{p}$, and ${\bf P}_{e}$
respectively. The joint probability amplitude distribution $F({\bf P}_{\nu },%
{\bf P}_{p},{\bf P}_{e})$ satisfies the normalization condition 
\begin{equation}
\int d^{3}{\bf P}_{\nu }d^{3}{\bf P}_{p}d^{3}{\bf P}_{e}\left\vert F({\bf P}%
_{\nu },{\bf P}_{p},{\bf P}_{e})\right\vert ^{2}=1.
\end{equation}%
The joint momentum of the entire antineutrino-proton-electron system is
essentially in a superposition of infinitely many components, which implies
that both the total momentum and energy are undeterministic. Despite these
uncertainties, the energy and momentum conservation laws are still satisfied
for each momentum component of the wave function, as correctly pointed out
in Ref. [9]. 

We here consider a specific component, denoted as $\left\vert {\bf P}_{\nu
}^{0},{\bf P}_{p}^{0},{\bf P}_{e}^{0}\right\rangle $. The momentum
conservation law implies that this momentum component originates from the
neutron momentum component $\left\vert {\bf P}_{n}^{0}\right\rangle $, with 
\begin{equation}
{\bf P}_{n}^{0}={\bf P}_{\nu }^{0}+{\bf P}_{p}^{0}+{\bf P}_{e}^{0}.
\end{equation}%
The energies of the neutron, proton, and electron associated with the
component $\left\vert {\bf P}_{\nu }^{0},{\bf P}_{p}^{0},{\bf P}%
_{e}^{0}\right\rangle $ are given by 
\begin{eqnarray}
E_{n}^{0} &=&\sqrt{m_{n}^{2}+(P_{n}^{0})^{2}},  \nonumber \\
E_{p}^{0} &=&\sqrt{m_{p}^{2}+(P_{p}^{0})^{2}},  \nonumber \\
E_{e}^{0} &=&\sqrt{m_{e}^{2}+(P_{e}^{0})^{2}},
\end{eqnarray}%
where $m_{n}$, $m_{p}$, and $m_{e}$ are the masses of the neutron, proton,
and electron, respectively. According to the energy conservation law, the
antineutrino's energy associated with the component $\left\vert {\bf P}_{\nu
}^{0},{\bf P}_{p}^{0},{\bf P}_{e}^{0}\right\rangle $ is definite, given by $%
E_{\nu }^{0}=E_{n}^{0}-E_{p}^{0}-E_{e}^{0}$. Consequently, the
antineutrino's mass is also definite, which is equal to $m_{\nu }=\sqrt{%
(E_{\nu }^{0})^{2}-(P_{\nu }^{0})^{2}}$. This leads to $m_{j}=m_{\nu }$ when 
$C_{j}\neq 0$. This result is inconsistent with the postulation that each
flavor eigenstate is a linear superposition of three different mass
eigenstates. If the flavor oscillations are caused by nonzero mass
differences, such a state cannot exhibit any oscillatory behavior.

{\sl lemma 2}: For the $\beta $ decay, different mass eigenstates of the
produced electron antineutrino are necessarily correlated with different
joint antineutrino-proton-electron momentum states.

Generally, the wavefunction of the entire system produced by the $\beta $
decay can be written in the form of

\begin{equation}
\left\vert \psi \right\rangle =%
\mathop{\displaystyle\sum}%
\limits_{j}\int_{{\bf \sigma }_{j}}d^{3}{\bf P}_{\nu ,j}d^{3}{\bf P}%
_{p,j}d^{3}{\bf P}_{e,j}G({\bf P}_{\nu ,j},{\bf P}_{p,j},{\bf P}%
_{e,j})\left\vert {\bf P}_{\nu ,j},{\bf P}_{p,j},{\bf P}_{e,j}\right\rangle
\left\vert \stackrel{-}{\nu }_{j}\right\rangle ,
\end{equation}%
where ${\bf \sigma }_{j}$ denotes the distribution region of the joint
antineutrino-proton-electron momentum associated with the antineutrino mass
eigenstate $\left\vert \stackrel{-}{\nu }_{j}\right\rangle $. In order to
satisfy the condition $m_{1}\neq m_{2}\neq m_{3}$, there should not be any
overlapping between the momentum distribution regions associated with
different mass eigenstates, i.e., ${\bf \sigma }_{j}\cap $ ${\bf \sigma }%
_{k}=\oslash $ for $j\neq k$. This can be interpreted as follows. Suppose
that there is an overlapping between the regions ${\bf \sigma }_{j}$ and $%
{\bf \sigma }_{k}$ with $j\neq k$. Then, according to the aforementioned
analysis, both $m_{j}$ and $m_{k}$ can be uniquely determined by a specific
joint momentum component $\left\vert {\bf P}_{\nu }^{0},{\bf P}_{p}^{0},{\bf %
P}_{e}^{0}\right\rangle $ that falls within the overlapping regime. This
implies $m_{j}=m_{k}$ when ${\bf \sigma }_{j}\cap $ ${\bf \sigma }_{k}\neq
\oslash $. 

For the entangled state of Eq. (8), when the momentum states are traced out,
the mass degree of freedom is left in a classical mixture, described by the
density operator 
\begin{eqnarray}
\rho _{\nu } &=&Tr_{{\bf P}_{\nu },{\bf P}_{n},{\bf P}_{e}}\left\vert \psi
\right\rangle \left\langle \psi \right\vert   \nonumber \\
&=&\int d^{3}{\bf P}_{\nu }d^{3}{\bf P}_{p}d^{3}{\bf P}_{e}\left\langle {\bf %
P}_{\nu },{\bf P}_{p},{\bf P}_{e}\right\vert \left. \psi \right\rangle
\left\langle \psi \right\vert \left. {\bf P}_{\nu },{\bf P}_{p},{\bf P}%
_{e}\right\rangle   \nonumber \\
&=&%
\mathop{\displaystyle\sum}%
\limits_{j,k}D_{j,k}\left\vert \stackrel{-}{\nu }_{j}\right\rangle
\left\langle \stackrel{-}{\nu }_{k}\right\vert ,
\end{eqnarray}%
with 
\begin{eqnarray}
D_{j,k} &=&\int_{{\bf \sigma }_{j}}d^{3}{\bf P}_{\nu ,j}d^{3}{\bf P}%
_{p,j}d^{3}{\bf P}_{e,j}\int_{{\bf \sigma }_{k}}d^{3}{\bf P}_{\nu ,k}d^{3}%
{\bf P}_{p,k}d^{3}{\bf P}_{e,k}  \nonumber \\
&&G({\bf P}_{\nu ,j},{\bf P}_{p,j},{\bf P}_{e,j})G^{\ast }({\bf P}_{\nu ,k},%
{\bf P}_{p,k},{\bf P}_{e,k})  \nonumber \\
&&\left\langle {\bf P}_{\nu ,k},{\bf P}_{p,k},{\bf P}_{e,k}\right\vert
\left. {\bf P}_{\nu ,j},{\bf P}_{p,j},{\bf P}_{e,j}\right\rangle .
\end{eqnarray}%
Since ${\bf \sigma }_{j}\cap $ ${\bf \sigma }_{k}=\oslash $ for $j\neq k$,
each of the joint momentum eigenstates in the region ${\bf \sigma }_{j}$ is
orthogonal to all the momentum eigenstates in ${\bf \sigma }_{k}$. This
implies $\left\langle {\bf P}_{\nu ,k},{\bf P}_{p,k},{\bf P}%
_{e,k}\right\vert \left. {\bf P}_{\nu ,j},{\bf P}_{p,j},{\bf P}%
_{e,j}\right\rangle =0$ throughout the integral region ${\bf \sigma }%
_{j}\otimes {\bf \sigma }_{k}$. Therefore, we have%
\begin{equation}
\rho _{\nu }=%
\mathop{\displaystyle\sum}%
\limits_{j}D_{j,j}\left\vert \stackrel{-}{\nu }_{j}\right\rangle
\left\langle \stackrel{-}{\nu }_{j}\right\vert ,
\end{equation}%
where%
\begin{equation}
D_{j,j}=\int_{{\bf \sigma }_{j}}d^{3}{\bf P}_{\nu ,j}d^{3}{\bf P}_{p,j}d^{3}%
{\bf P}_{e,j}\left\vert G({\bf P}_{\nu ,j},{\bf P}_{p,j},{\bf P}%
_{e,j})\right\vert ^{2}.
\end{equation}%
In other words, the quantum coherence among the mass eigenstates is
destroyed by their quantum entanglement with different joint momentum states.

The entanglement-induced loss of coherence can also be understood in terms
of complementarity, according to which the interference between state
components of one freedom degree would be destroyed if the information about
it is stored in another freedom degree [15-24]. Here the information about
the mass of the antineutrino is encoded in the joint
antineutrino-proton-electron momentum. This information could be extracted
in principle, which is sufficient to destroy the interference between the
mass eigenstates. It does not matter whether or not the information is read
out. The loss of interference due to entanglement has been demonstrated in a
number of experiments [18-24].

If antineutrino flavor eigenstates are defined as superpositions of mass
eigenstates, for such a classical mixture $\rho _{\nu }$, the population of
the flavor eigenstate $\left\vert \stackrel{-}{\nu }_{\alpha }\right\rangle $
is given by 
\[
P_{\alpha }=%
\mathop{\displaystyle\sum}%
\limits_{j}\left\vert D_{j,j}U_{j\alpha }\right\vert ^{2}, 
\]%
where 
\[
\left\vert \stackrel{-}{\nu }_{j}\right\rangle =%
\mathop{\displaystyle\sum}%
\limits_{j}U_{j\alpha }^{\dagger }\left\vert \stackrel{-}{\nu }_{\alpha
}\right\rangle . 
\]%
Consequently, the probability for detecting the antineutrino in each flavor
has a nonzero probability, which is time-independent. This is inconsistent
with the well-known $\beta $ decay experiments [4], where the produced
antineutrino is initially of e-type, and then undergoes flavor oscillations.

In summary, we have shown that the electron antineutrino created by the $%
\beta $ decay of a neutron cannot be in a coherent superposition of
different mass eigenstates. If the entire antineutrino-proton-electron
wavefunction emerging from the $\beta $ decay involve different mass
eigenstates, they are necessarily entangled with different momentum states,
as a consequence of the momentum and energy conservation laws. Such
mass-momentum entanglement would destroy the quantum coherence between the
mass eigenstates. This conclusion holds for any neutrino or antineutrino
produced by the decay of an unstable particle. The result enforces the claim
that neutrino oscillations originate from virtual excitation of the Z
bosonic field [25].

\end{document}